\documentclass{llncs}

\usepackage{enumitem}
\usepackage{comment}
\usepackage{cite}
\usepackage{xspace}
\usepackage{commands}
\usepackage{language}
\usepackage{calc}
\usepackage{multirow}
\usepackage{amsmath,amssymb,latexsym}
\usepackage{amsfonts}
\usepackage{mathtools}
\usepackage{booktabs}

\usepackage{comment}
\usepackage{xspace}
\usepackage{hyperref}
\usepackage[inference]{semantic}
\usepackage{lmodern}

\usepackage{language}
\usepackage{commands}
\usepackage[numbers]{natbib}

\usepackage{showexpl}% includes listings
\usepackage{amssymb}

\def\codesize{\normalsize}

\definecolor{rubystr}{gray}{0.1}
\definecolor{gray_ulisses}{gray}{0.5}
\definecolor{preto_ulisses}{gray}{0.0}

\lstdefinelanguage{ruby}{
	basicstyle=\sf\fontseries{m}\selectfont\codesize,
        columns=flexible,
	sensitive=true,
	aboveskip=\smallskipamount, 
	belowskip=\smallskipamount, 
	morecomment=[l][\color{gray_ulisses}\sf\fontseries{m}\selectfont\itshape\codesize]{\#\#},
	morestring=[b]",
    escapeinside={(*}{*)},
	stringstyle=\color{rubystr},
	showstringspaces=false,
	numberstyle=\codesize,
	numberblanklines=true,
	morestring=[b]",
	showspaces=false,
	breaklines=true,
	showtabs=false,
	literate={ {->}{{$\rightarrow$}}2
	  		   {<=}{{$\leq$}}2
	  		   {&&}{{$\land$}}2
	  		   {=>}{{$\Rightarrow$}}2
             },
    	emph=
	{[1]
	    class, def, if, then, else, end, return, label, in, for,
	    module, Folder, attr_accessor, initialize, include,
	    var_type, pure, protects, modifies,
	    int, bool, Integer, Int, Bool,
	    pre, type, verify, struct, object, self, exact, set,
	    Time, Arithmetic, UserFile,  ActiveRecord, Base, ActiveRecord, Associations, ClassMethods, Money, 
	    Float, Array,
	    rdl_do_verify, 
	    assume, assert
	},
	emphstyle={[1]\sf\fontseries{b}\selectfont}
}

\lstdefinelanguage{rosette}{
	basicstyle=\sf\fontseries{m}\selectfont\codesize,
        columns=flexible,
	sensitive=true,
	aboveskip=\smallskipamount, 
	belowskip=\smallskipamount, 
	morecomment=[l][\color{gray_ulisses}\sf\fontseries{m}\selectfont\textit\codesize]{\#\#},
	morestring=[b]",
    escapeinside={(*}{*)},
	stringstyle=\color{rubystr},
	showstringspaces=false,
	numberstyle=\codesize,
	numberblanklines=true,
	showspaces=false,
	breaklines=true,
	showtabs=false,
	literate={ {->}{{$\rightarrow$}}1
	  		   {<=}{{$\leq$}}2
	  		   {&&}{{$\land$}}1
             },
    	emph=
	{[1]
	    define, verify, assume, assert, guarantee, let, in, struct, object, mutable, 
	    if, then, else, set, symbolic
	},
	emphstyle={[1]\sf\fontseries{b}\selectfont}
}

\lstdefinestyle{rosette}{basicstyle=\sf\small,language=Lisp,numbers=left,numberstyle=\tiny,showstringspaces=false,escapeinside={(*}{*)},breaklines=true,keywords={}}

\lstnewenvironment{rscode}
{\lstset{language=rosette}}
{}

\lstnewenvironment{rcode}
{\lstset{language=ruby}}
{}

\lstMakeShortInline[language=ruby,mathescape,basicstyle=\sf\fontseries{m}\selectfont\normalsize,breakatwhitespace,xleftmargin=0pt,xrightmargin=0pt]^

\DeclareTextFontCommand{\code}{\sf}

\begin{document}

\title{Refinement Types for Ruby}

%%% All paper formatting guidelines are here: ftp://ftp.springernature.com/cs-proceeding/svproc/guidelines/Springer_Guidelines_for_Authors_of_Proceedings.pdf

\author{Milod Kazerounian\inst{1} \and Niki Vazou\inst{1} \and Austin
  Bourgerie\inst{1} \and Jeffrey\,S.\,Foster\inst{1} \and Emina Torlak\inst{2}}
\authorrunning{Milod Kazerounian et al.}

\institute{University of Maryland, College Park, USA\\
\email{\{milod, nvazou, abourg, jfoster\}@cs.umd.edu},\\ 
\and University of Washington, Seattle, USA\\
\email{emina@cs.washington.edu}}

\maketitle

\begin{abstract}

  Refinement types are a popular way to specify and reason about key
  program properties. In this paper, we introduce RTR, a new system
  that adds refinement types to Ruby. RTR is built on top of RDL, a
  Ruby type checker that provides basic type information for the
  verification process. RTR works by encoding its verification
  problems into Rosette, a solver-aided host language. RTR handles
  mixins through assume-guarantee reasoning and uses just-in-time
  verification for metaprogramming. We formalize RTR by showing a
  translation from a core, Ruby-like language with refinement types
  into Rosette. We apply RTR to check a range of functional
  correctness properties on six Ruby programs. We find that RTR can
  successfully verify key methods in these programs, taking only a few
  minutes to perform verification.

\keywords{Ruby, Rosette, refinement types, dynamic languages}
\end{abstract}

%%\textit{Refinement types} give programmers the ability to write and check 
%%expressive specifications over programs. 
%%But dynamic language features, such as \textit{mixins} which allow for partial environments and \textit{metaprogramming}
%%which allows for the dynamic generation and execution of code, pose tough challenges
%%to verifying refinement types. In this paper, we extend the approach of 
%%\textit{just-in-time static type checking} to the verification of 
%%refinement types and present \rrdl, a refinement type checker for the highly
%%dynamic language \ruby. To achieve verification, we translate \ruby programs
%%to \rosette, a language with constructs for verification. We formalize this translation
%%using a core subset of \ruby, and we evaluate \rrdl on a number of \ruby
%%libraries and applications.

\section{Introduction}\label{sec:intro}
Refinement types combine types with logical predicates to 
encode program invariants~\cite{Xi98,Rushby98}. For example, the
following refinement type specification:
\begin{rcode}
  type :incr_sec, `(Integer x { 0 <= x < 60 }) -> Integer r { 0 <= r < 60}'
\end{rcode}
describes a method ^incr_sec^ that increments a second. With this
specification, ^incr_sec^ can only be called with integers that are
valid seconds (between 0 and 59) and the method will always return
valid seconds.

Refinement types were introduced to reason about simple invariants,
like safe array indexing~\cite{Xi98}, but since then they have been
successfully used to verify sophisticated properties including
termination~\cite{Vazou14}, program equivalence~\cite{Aguirre17}, and
correctness of cryptographic protocols~\cite{Protzenko17}, in various
languages (\eg ML~\cite{Freeman91}, Racket~\cite{Kent16}, and
TypeScript~\cite{Vekris16}).

In this paper, we explore refinement types for Ruby, a popular,
object-oriented, dynamic scripting language. Our starting place is
RDL~\cite{Ren, rdl-github}, a Ruby type system recently developed by
one of the authors and his collaborators. We introduce \rrdl{}, a tool
that adds refinement types to RDL and verifies them via a translation
into Rosette~\cite{Torlak13}, a solver-aided host language. Since
Rosette is not object-oriented, \rrdl{} encodes Ruby objects as
Rosette structs that store object fields and an integer identifying
the object's class. At method
calls, \rrdl{} uses RDL's type information to statically overestimate
the possible callees. When methods
with refinement types are called,
\rrdl{} can either translate the callee directly or treat it modularly by asserting the method preconditions and assuming the postcondition,
using purity annotations to determine which fields (if any) the method
may mutate. (\S~\ref{sec:overview})

In addition to standard object-oriented features, Ruby includes
dynamic language features that increase flexibility and
expressiveness. In practice, this introduces two key challenges in
refinement type verification: \emph{mixins}, which are Ruby code
modules that extend other classes without direct inheritance,
and \emph{metaprogramming}, in which code is generated on-the-fly
during runtime and used later during execution. The latter feature is
particularly common in Ruby on Rails, a
popular Ruby web development framework.

To meet these challenges, \rrdl{} uses two key ideas. First, \rrdl{}
incorporates \emph{assume-guarantee checking}~\cite{Jones83} to reason
about mixins. \rrdl{} verifies definitions of methods in mixins by
assuming refinement type specifications for all undefined, external
methods. Then, by dynamically intercepting the call that includes a
mixin in a class, \rrdl{} verifies the appropriate class methods
satisfy the assumed refinement types (\S~\ref{subsec:mixins}). Second,
\rrdl{} uses \emph{just-in-time verification} to reason about
metaprogramming, following RDL's just-in-time type
checking~\cite{Ren}. In this approach, (refinement) types are
maintained at run-time, and methods are checked against their types
after metaprogramming code has executed but before the methods have
been called (\S~\ref{subsec:metaprogramming}).

We formalized \rrdl{} by showing how to translate \corelan, a core
Ruby-like language with refinement types, into \coreinter, a core
verification-oriented language. We then discuss how to map the latter
into \rosette{}, which simply requires encoding \coreinter{}'s
primitive object construct into \rosette{} structs and translating
some control-flow constructs such as \code{return}
%We define the translation of \corelan to \rosette and claim that if
%all labels are sound and all refinements are pure, \rrdl is also sound
(\S~\ref{sec:formalism}).

We evaluated \rrdl{} by using it to check a range of functional
correctness properties on six Ruby and Rails applications. 
%%In \rrdl{},
%%verification is done on a per-method basis, and thus verified and
%%unverified methods can co-exist and depend on each other, allowing for
%%gradual verification. 
In total, we verified 31 methods,
comprising 271 lines of Ruby, by encoding them as 1,061 lines of
Rosette. We needed 73 type annotations. Verification took a total median
time (over multiple trials) of 506 seconds
(\S~\ref{sec:evaluation}).

Thus, we believe \rrdl{} is a promising first step toward
verification for Ruby.

% Finally, we implemented our technique by adding logical refinements to
% the types of \rdl~\cite{Ren}, a just-in-type type checker for \ruby
% programs.  We used the implementation to check functional correctness,
% array and numeric invariants on six real \ruby and \rails applications
% (\S~\ref{sec:evaluation}).  In \rrdl verification is called per method
% (instead of per module), thus verified and unverified methods can
% co-exist and depend on each other, allowing for gradual verification.
% This, combined with the reasonable amount of the required annotations
% (we used ^64^ specifications to verify ^214^ lines of \ruby), makes
% \rrdl a usable \ruby verifier.  \NV{Assume array bench is there}
% \NV{Update annotation burden if we add array}

\section{Overview}\label{sec:overview}

We start with an overview of \rrdl, 
which extends the Ruby type checker \rdl{}~\cite{Ren} with refinement types.
In \rrdl, program invariants are specified with refinement types (\S~\ref{subsec:specs}) 
and checked by translation to \rosette (\S~\ref{subsec:overview:rosette}).
We  translate
\ruby objects to \rosette structs (\S~\ref{subsec:objects})
and method calls to function calls (\S~\ref{subsec:method-calls}).

\subsection{Refinement Type Specifications}\label{subsec:specs}

Refinement types in \rrdl 
are \ruby types extended with logical predicates. For example,
we can use \rdl's ^type^ method to link a method with its
specification:
\begin{rcode}
  type `(Integer x { 0 <= x < 60 }) -> Integer r { 0 <= r < 60}'
  def incr_sec(x) if (x==59) then 0 else x+1 end ; end
\end{rcode}
This type indicates the argument and result of ^incr_sec^ are integers
in the range from ^0^ to ^59^. In general, refinements (in curly
braces) may be arbitrary Ruby expressions that are treated as
booleans, and they should be \textit{pure}, \ie have no side effects,
since effectful predicates make verification either complicated or
imprecise~\cite{halo}. As in \rdl{}, the type annotation, which is a
string, is parsed and stored in a global table which maintains the
program's type environment.

\subsection{Verification using \rosette}\label{subsec:overview:rosette}
\rrdl{} checks method specifications by encoding their verification
into \rosette~\cite{Torlak13}, a solver-aided host language
built on top of Racket. Among other features, Rosette can perform
verification by using symbolic execution to generate logical
constraints, which are discharged using Z3~\cite{deMoura}.

For example, to check ^incr_sec^, \rrdl{} creates the 
equivalent \rosette{} program:
\begin{rscode}
  (define (incr_sec x) (if (= x 59) 0 (+ x 1)))
  (define-symbolic x_in integer?)
  (verify #:assume (assert 0 <= x < 60)
          #:guarantee (assert (let ([r (incr_sec x)]) 0 <= r < 60)))
\end{rscode}
Here ^x_in^ is an integer \textit{symbolic constant} representing
an unknown, arbitrary integer argument. \rosette symbolic
constants can range over the \textit{solvable types} integers,
booleans, bitvectors, reals, and uninterpreted functions. We use \rosette's
^verify^ function with assumptions and assertions to encode pre- and
postconditions, respectively.
When this program is run, Rosette searches for an ^x_in^ such that
the assertion fails. If no such value exists, then the assertion
is verified.

%\rosette also includes a number of useful features from
%Racket, including structs and datatypes such as lists and vectors,
%that we will see below.
% We bind the method's result to
%a variable ^r^ to avoid multiple computations.

\subsection{Encoding and Reasoning about Objects}\label{subsec:objects}

We encode \ruby objects in \rosette using a \emph{struct type}, i.e.,
a record. More specifically, we create a struct type ^object^
that contains
an integer ^classid^ identifying the object's class, 
an integer ^objectid^ identifying the object itself, and 
a field for each instance variable of all objects encountered 
in the source \ruby program (similarly to prior work~\cite{Jeon,Singh}).

For example, consider a \ruby class ^Time^ 
with three instance variables ^@sec^, ^@min^, and ^@hour^,
and a method ^is_valid^ that checks all three variables are valid:
\begin{rcode}
  class Time
    attr_accessor :sec, :min, :hour
     
    def initialize(s, m, h) @sec = s; @min = m; @hour = h; end

    type `() -> bool'
    def is_valid 0 <= @sec < 60 && 0 <= @min < 60 && 0 <= @hour < 24; end
  end
\end{rcode}
\rrdl{} observes three fields in this program, and thus it defines:
\begin{rscode}
  (struct object ([classid][objectid]
                  [@sec  #:mutable] [@min  #:mutable] [@hour #:mutable]))
\end{rscode}
Here ^object^ includes fields for the class ID,
object ID, and the three instance variables.  Note since
^object^'s fields are statically defined, our encoding cannot
handle dynamically generated instance variables, which we leave as
future work.

%JF: No real need to say the following here...
% in \rosette all \ruby objects, apart from the primitive ones.  We
% encode primitive objects, like booleans, numeric types, and arrays to
% their respective primitive \rosette datatypes, to achieve precise
% reasoning over these theories (\S~\ref{subsec:primitives}).

RTR then translates Ruby field reads or writes as
getting or setting, respectively, ^object^'s fields in Rosette. For
example, suppose we add a method ^mix^ to the ^Time^ class and specify
it is only called with and returns valid times:
\begin{rcode}
  type :mix, `(Time t1 { t1.is_valid},  Time t2 { t2.is_valid},
               Time t3 { t3.is_valid})  -> Time r { r.is_valid}'
  def mix(t1,t2,t3) @sec = t1.sec; @min = t2.min; @hour = t3.hour; self; end
\end{rcode}
Initially, type checking fails because the getters' and setters' (\eg
^sec^ and ^sec=^) types are unknown. Thus, we add those types:
\begin{rcode}
  type :sec,  `()          -> Integer i   { i == @sec }'
  type :sec=, `(Integer i) -> Integer out { i == @sec }'
\end{rcode}
(Note these annotations can be generated automatically using our
approach to metaprogramming, described in \S~\ref{subsec:metaprogramming}.) This
allows RTR to proceed to the translation stage, which generates the
following \rosette function:
\begin{rscode}
(define (mix self t1 t2 t3)
   (set-object-@sec!  self  (sec  t1))
   (set-object-@min!  self  (min  t2))
   (set-object-@hour! self  (hour t3))
   self)
\end{rscode}
(Asserts, assumes, and verify call omitted.)
Here ^(set-object-x! y w)^
writes ^w^ to the ^x^ field of ^y^
and the field selectors ^sec^, ^min^, and ^hour^
are uninterpreted functions. 
Note that ^self^
turns into an explicit additional argument in the \rosette definition.
% Verification of ^mix^ reduces to a \rosette ^verify^ query 
%that, assuming ^mix^'s arguments are valid, 
%asserts that the result is valid as well. 
%
\rosette{} then verifies this program, thus verifying the original
\ruby{} ^mix^ method.

\subsection{Method Calls}\label{subsec:method-calls}

To translate a \ruby method call ^e.m(e1, .., en)^, \rrdl{} needs to
know the callee, which depends on the runtime type of the receiver
^e^. \rrdl{} uses RDL's type information to overapproximate the set of
possible receivers. For example, if ^e^ has type ^A^ in RDL, then
\rrdl{} translates the above as a call to ^A.m^. If ^e^ has a
union type, \rrdl{} emits Rosette code that branches on the potential
types of the receiver using ^object^ class IDs and dispatches to the
appropriate method in each branch. This is similar to class
hierarchy analysis~\cite{Dean}, which also uses types to determine the
set of possible method receivers and construct a call graph. 
%Because type checking
%is necessary to method call translation, we require the programmer to provide
%type annotations for \textit{all} methods and instance variables to be translated
%to Rosette.

% The translation resolves the type of the receiver ^e^ by asking \rdl,
% the just-in-type unrefined-type checker.  To type an expression ^e^,
% \rdl requires the user to provide type annotations for any methods and
% instance variables that appear inside ^e^ and at runtime tags each
% subexpression of ^e^ with its type (similar to~\citet{Dean}).
% %
% This requirement works well under metaprogramming, 
% since the user can also use metaprogramming 
% to provide a type for each dynamically defined method (\S~\ref{subsec:metaprogramming}).
% %
% To get the type of ^e^, the translation simply look up its tag. 
% %

% To translate the method call ^e.m(e1, .., en)^ in \rosette, we first
% use \rdl to find the class (say ^A^) that the receiver belongs to and
% then correctly dispatch the call to the method ^A.m^.  Should a
% receiver have a union type, we emit \rosette code which branches on
% the potential types of the receiver using ^object^ class IDs, and
% dispatches the appropriate method in each branch.

Once the method being called is determined, we translate the call into
\rosette.  As an example, consider a method ^to_sec^ that converts
^Time^ to seconds, after it calls the method ^incr_sec^
from~\S~\ref{subsec:specs}.
\begin{rcode}
  type `(Time t { t.is_valid}) -> Integer r { 0<=r<90060 }'
  def to_sec(t) incr_sec(t.sec) + 60 * t.min + 3600 * t.hour; end
\end{rcode}
%The specification of ^to_sec^ requires the result to be less than
%^90060^, which must be true when the ^hour^, ^min^ and increased ^sec^
%are valid.

\rrdl{}'s translation of ^to_sec^ could simply call directly into ^incr_sec^'s
translation. This is equivalent to inlining ^incr_sec^'s code.
However, inlining is not always possible or desirable. A method's code
may not be available because the method comes from a library, 
is external to the environment (\S~\ref{subsec:mixins}), or has not been defined yet (\S~\ref{subsec:metaprogramming}). The method might also contain constructs 
that are difficult to verify, like diverging loops. 

Instead, RTR can model the method call
using the programmer provided method specification.
%For verification, one could just inline the definition of ^incr_sec^, 
%but this is not always possible---the method code may not be available
%(e.g., if it is a library method)---or desirable---e.g., the method
%code may be large and complex.
%To translate such calls,  
%\rrdl{} uses assume-guarantee reasoning.
To precisely reason with only a method's specification, RTR follows
Dafny~\cite{Leino10} and treats pure and impure methods differently.

\paragraph{Pure methods.}
Pure methods 
have no side effects
and return the same result for the same inputs, 
satisfying the congruence property $\forall x, y. x = y \Rightarrow m(x) = m(y)$
for a given method $m$.
Thus, pure methods can be encoded using \rosette's uninterpreted functions.
The method ^incr_sec^ is indeed pure,
so we can label it as such:
\begin{rcode}
  type :incr_sec, `(Integer x { 0<=x<60 }) -> Integer r { 0<=r<60 }', :pure
\end{rcode}
With the ^pure^ label, the translation of 
^to_sec^ treats ^incr_sec^ as an uninterpreted function.
Furthermore, it asserts the precondition ^0<=x<60^
and assumes the postcondition ^0<=r<60^, 
which is enough to verify ^to_sec^.

\paragraph{Impure methods.}
Most \ruby methods have side effects and
thus are not pure.
For example, consider ^incr_min^, a mutator method that adds a
minute to a ^Time^:
\begin{rcode}
  type `(Time t { t.is_valid && t.min < 59 }) -> Time r { r.is_valid }',
    modifies: { t: @min, t: @sec }
  def incr_min(t)
    if t.sec<59 then t.sec=incr_sec(t.sec) else t.min+=1; t.sec=0 end
    return t
  end
\end{rcode}
A translated call to ^incr_min^ generates a fresh symbolic value as the
method's output and assumes the method's postcondition on that value.
Because the method may have side effects, the ^modifies^ label is used
to list all fields of inputs which may be modified by the method.
Here, a translated call to ^incr_min^ will \textit{havoc} (set to fresh symbolic values)
^t^'s ^@min^ and ^@sec^ fields.

We leave support for other kinds of modifications
(e.g., global variables, transitively reachable fields), as well as enforcing
the ^pure^ and ^modifies^ labels, as future work.

% In summary, our implementation provides three labels on method
% specifications, each of which uses a different rule in our translation
% (Figure~\ref{fig:translation}).
% The ^exact^ label (Rule~\rtnonmodcall) inlines the method definition
% (thus, can can be used only when the definition is available), the
% ^pure^ label (Rule~\rtmodpurecall) marks effect-free, congruent
% methods, and the ^protects {ti => fi}^ label (Rule~\rtmodimpurecall)
% protects the field ^fi^ of the input ^ti^.
% The default (and sound) label is protecting nothing ^protects {}^. 
% %
% Currently, the labels are trusted and checking them
% is left as future work. 

\section{Just-In-Time Verification}\label{sec:read-ruby}
Next, we show how RTR handles code with 
dynamic bindings via mixins (\S~\ref{subsec:mixins}) 
and metaprogramming (\S~\ref{subsec:metaprogramming}).
%
%In both cases, verification happens just-in-time: 
%verification of a method cannot happen statically, 
%since the method's environment is not yet defined, 
%but happens \textit{before} the method is executed, 
%thus ensuring that no violations occur at runtime. 

\subsection{Mixins}\label{subsec:mixins}
Ruby implements mixins via its \textit{module} system.
A Ruby module is a collection of method definitions 
that are added to any class that ^include^s the module at runtime.
Interestingly, modules may refer to methods that are not defined in
the module but will ultimately be defined in the including class.
Such incomplete environments pose a challenge for verification. 

Consider the following method that has been extracted and simplified
from the \money library described in \S~\ref{sec:evaluation}. 
\begin{rcode}
  module Arithmetic
    type `(Integer x)-> Float r { r==x/value }'
    def div_by_val(x) x/value; end
  end
\end{rcode}
The module method ^div_by_val^ divides its input ^x^ by ^value^.
\rrdl's specification for ^/^ requires 
that ^value^ cannot be ^0^.
%
%The above lack of division-by-zero example
%is elementary for standard refinement types, 
%yet new light is shed in the dynamic contexts of \ruby.
 
Notice that ^value^ is not defined in ^Arithmetic^. Rather, it must be
defined wherever ^Arithmetic^ is included.
Therefore, to proceed with verification in RTR, the programmer must provide an annotation for ^value^:
\begin{rcode} % [style=ruby,numbers=none]
  type :value, `() -> Float v { 0 < v }', :pure
\end{rcode}
Using this annotation, RTR can verify ^div_by_value^.
%
% Since ^value^ is marked as pure, as discussed above RTR encodes it as
% an uninterpreted function, thereby
% allowing verification of ^div_by_val^ to succeed. 
% %
% Note that if ^value^ were impure,  verification would be impossible, 
% since it relies on congruence of ^value^.
%%since it relies on different calls to ^value^ (in the method
%%body and in the type refinement) with
%%the same input returning the same output.
%
Then when ^Arithmetic^ is included in another class, RTR verifies ^value^'s
refinement type. For example, consider the following code:

\begin{rcode}
  class Money
    include Arithmetic
    def value() 
      if (@val > 0) then (return @val) else (return 0.01) end
    end
  end
\end{rcode}

RTR dynamically intercepts the call to ^include^ and then applies the
type annotations for methods not defined in the included module, in
this case verifying ^value^ against the annotation in ^Arithmetic^.
Thus, RTR follows an assume-guarantee paradigm \cite{Jones83}: it
assumes ^value^'s annotation while verifying ^div_by_val^ and then
guarantees the annotation once ^value^ is defined.

%% \NV{Getting rid of implementation, too low level....}
%This is implemented by maintaining a list for each module of annotations which should be applied wherever the module is included. We then monkey patch Ruby's ^name=^include} method so that it not only includes a module, but also applies this list of annotations to the place where the modules is included. If it is included in a class, this list of annotations will be applied to methods within that class (or will wait for new methods to be defined within that class). If it is included in another module, then the annotations will be applied both within the module, and will be added to that module's list of annotations to be applied externally.

\subsection{Metaprogramming}\label{subsec:metaprogramming}

Metaprogramming in \ruby allows new methods to be created and
existing methods to be redefined on the fly, posing a challenge for
verification. RTR addresses this challenge using just-in-time
checking~\cite{Ren}, in which, in addition to code, method annotations
can also be generated dynamically.

We illustrate the just-in-time approach 
using an example from \boxroom{}, a \rails app for
managing and sharing files in a web browser (\S~\ref{sec:evaluation}).
The app defines a class ^UserFile^ that is 
a Rails \textit{model} corresponding to a database table:
\begin{rcode} % [style=ruby]
  class UserFile < ActiveRecord::Base
    belongs_to :folder
    ...type `(Folder target) -> Bool b { folder == target }'
       def move(target) folder = target; save!; end...
  end
\end{rcode}
Here calling ^belongs_to^ tells Rails that every ^UserFile^ is
associated with a ^folder^ (another model).  The ^move^ method updates
the associated folder of a ^UserFile^ and saves the result to the
database.  We annotate ^move^ to specify that the ^UserFile^'s new
folder should be the same as ^move^'s argument.

This method and its annotation are seemingly simple, but there is a
problem. To verify ^move^, RTR needs an annotation for the ^folder=^
method, which is not statically defined. Rather, it is dynamically
generated by the call to ^belongs_to^.

To solve this problem in RTR, we instrument ^belongs_to^ to generate
type annotations for the setter (and getter) method, as follows:
\begin{rcode} % [style=ruby]
  module ActiveRecord::Associations::ClassMethods
    pre(:belongs_to) do |*args|
      name  = args[0].to_s
      cname = name.camelize
      type `#{name}' ,  `() -> #{cname} c', :pure
      type `#{name}=', `(#{cname} i) -> #{cname} o { #{name} == i }'
      true
    end
  end
\end{rcode}
%
%           protects: []
%\NV{remove the protects label above as default}
We call ^pre^, an \rdl method, to define a precondition \emph{code
  block} (\ie an anonymous function) which will be executed on each
call to ^belongs_to^.
First, the block sets
^name^ and ^cname^ to be 
the string version of the first argument passed to ^belongs_to^
and its camelized representation, respectively. 
%
%In our example ^belongs_to :folder^ call, ^name^ and ^cname^ will be set to 
%^`folder'^ and \textsf{`Folder'}, respectively. 
%
Then, we create types for the ^name^ and ^name=^ methods. Finally, we
return ^true^ so the contract will succeed.
In our example, this code generates the following two type annotations
upon the call to ^belongs_to^:
\begin{rcode} % [style=ruby]
  type `folder' , `() -> Folder c', :pure
  type `folder=', `(Folder i) -> Folder o { folder == i }'
\end{rcode}
%
%, protects: []
%
%These annotations will be generated 
%when ^belongs_to^ is invoked with the ^:folder^ argument, 
%which happens exactly after the ^UserFile^ class is loaded.
%
%Thus, not only will the call to ^belongs_to^ generate getter
%and setter methods, but it also generates useful annotations
%for these methods.
With these annotations,
verification of the initial ^move^ method succeeds.
%
%Even though ^folder=^ is not labeled as pure
%(and thus does not satisfy congruence), 
%its postcondition exactly captures the 
%verification requirement of ^move^
%that ^folder == target^.  

%In sum, with just-in-time specification we can
%verify methods in the face of metaprogramming.
%
%We use dynamically provided information to generate expressive annotations 
%at runtime for methods that do not even exist statically.

\section{From Ruby to Rosette}\label{sec:formalism}

In this section, we formally describe our verifier and the translation
from \ruby to \rosette.
We start (\S~\ref{subsec:syntax}) by defining \corelan, a \ruby subset
that is extended with refinement type specifications. 
We give (\S~\ref{subsec:tranlation}) a translation from \corelan to an
intermediate language \coreinter,
and then (\S~\ref{subsec:rosette}) we discuss how \coreinter 
maps to a \rosette program. 
Finally (\S~\ref{subsec:soundness}), 
we use this translation to construct a verifier for \ruby.

\subsection{Core \ruby \corelan and Intermediate Representation \coreinter}\label{subsec:syntax}
\begin{figure}[t!]
\centering
$$
\begin{array}{rrcl}
\emphbf{Constants} \quad
  & c
  & ::=& \enil \spmid \etrue \spmid \efalse \spmid 0, 1,-1, \dots
\\[0.03in]

\emphbf{Expressions} \quad
  & \expr
  & ::=& c \spmid x \spmid \eass{x}{\expr} \spmid \eif{\expr}{\expr}{\expr} \spmid \eseq{\expr}{\expr}\\
  &   & \spmid & \eself % \spmid \enew{\tclass} 
        \spmid \einst{\field} \spmid \eass{\field}{\expr} 
        \spmid \ecall{\expr}{m}{\overline{\expr}} \spmid \enew{\tclass}\spmid\ereturn{\expr}
\\[0.03in]

\emphbf{Refined Types} \quad
  & \rtyp
  & ::=& \tref{x}{\tclass}{\expr}
\\[0.03in]

\emphbf{Program} \quad
  & \pg & ::=& \cdot \spmid d, \pg \spmid a, \pg
\\[0.03in]

\emphbf{Definition} \quad
  & d & ::=& 
  \pgdef{\methname{\tclass}{m}}
        {\overline{{\rtyp}}}
        {\rtyp}
        {\tlabel}
        {\expr}
\\[0.03in]

\emphbf{Annotation} \quad
  & a & ::=& \typdef{\methname{\tclass}{m}}{\tfun{\overline{{\rtyp}}}{\rtyp}}{\tlabel}
  \quad\quad\quad\quad\quad\quad \text{with}\ \tlabel \not = \lexact
\\[0.03in]

\emphbf{Labels} \quad
  & \tlabel
  & ::=& \lexact \spmid \lpure \spmid \lmodifies{\overline{\efieldread{\field}{x}}}
\end{array}
$$

$x\in\textrm{var ids}$, $\field\in\textrm{field ids}$, $m\in\textrm{meth ids}$, $A\in\textrm{class ids}$

\caption{\textbf{Syntax of the Ruby Subset \corelan}.}
\label{fig:ruby-syntax}
\end{figure}

\begin{figure}[t!]
\centering
$$
\begin{array}{rrcl}
\emphbf{Values} \quad
  & \ival
  & ::=& c \spmid \vobject{i}{i}{\overline{\ibind{\field}{\ival}}}
\\[0.03in]
\emphbf{Expressions} \quad
  & \iexpr
  & ::=& \ival \spmid x \spmid \eass{x}{\iexpr} \spmid \eif{\iexpr}{\iexpr}{\iexpr} \spmid \eseq{\iexpr}{\iexpr}\\
  &   & \spmid & \elet{x}{\overline{\ibind{x}{\iexpr}}}{\iexpr} 
        \spmid \efunc{x}{\overline{\iexpr}} \spmid \eassert{\iexpr}\\
  &   & \spmid & \eassume{\iexpr} \spmid \ereturn{\iexpr} \spmid \ehavoc{\efieldread{f}{x}} \spmid \efieldass{x}{\field}{\iexpr} \spmid \efieldread{\field}{x}
\\[0.03in]

\emphbf{Program} \quad
  & \ipg
  & ::=& \cdot \spmid d, \ipg \spmid v, \ipg
\\[0.03in]

\emphbf{Definition} \quad
  & d
  & ::=& \define{x}{\overline{x}}{\iexpr} \spmid \definesymobj{x}{\ityp}
\\[0.03in]

\emphbf{Verification Query} \quad
  & v
  & ::=& \verify{\overline{\iexpr}}{\iexpr}
\\[0.03in]

%\emphbf{Symbolic Types} \quad
%  & \ityp
%  & ::=& \symint \spmid \symbool \spmid \symreal \spmid \symfun{\overline{\ityp}}{\ityp}
%\\[0.03in]
\end{array}
$$

$x\in\textrm{var ids}$, $\field\in\textrm{field ids}$, $\ityp\in \textrm{types}$, $i\in\textrm{integers}$
\caption{\textbf{Syntax of the Intermediate Language \coreinter}.}
\label{fig:inter-syntax}
\end{figure}
%%
%%\NV{Can symbolic type be higher order? That is with function argument?}
%%\NV{Why did you remove symbolic types???}

%
\subsubsection{\corelan.}
Figure~\ref{fig:ruby-syntax} defines \corelan, a core Ruby-like language
with refinement types.
\textit{Constants} consist of \enil, booleans, and integers.
\textit{Expressions} include constants, variables, assignment, conditionals, sequences, and the reserved variable \eself, which refers to a method's receiver.
Also included are references to an instance variable \einst{f} and
instance variable assignment; we note that in actual \ruby, field names
are preceded by a ``@''. 
Finally, expressions include method calls, constructor calls 
\enew{\tclass} which create a new instance of class \tclass,
and return statements.
%
%%To create a new instance of class \tclass we do not need special syntax, 
%%instead we call a method \texttt{new}, \ie $\ecall{\tclass}{\texttt{new}}{}$.
%%\NV{This does not make sense, since A is not an expression.}
%%\MK{I am not sure this last sentence is necessary as I think it confuses more than clarifies. But basically, it is still okay because A is actually an instance of the Class class, so it is an [A] and therefore it is an expression.}

\textit{Refined types} \tref{x}{\tclass}{\expr}
refine the basic type \tclass with the predicate \expr.
The basic type \tclass 
is used to represent 
both user defined  and built-in classes
including nil, booleans, integers, floats, etc.
The refinement \expr is 
a \textit{pure, boolean valued} expression that may refer to the refinement variable  
$x$.
%
% That is, the variable $x$ can appear free in the expression \expr. 
%
In the interest of greater simplicity for the translation,
we require that \eself \textit{does not} appear in refinements \expr; 
however, extending the translation to handle this is natural, and our implementation allows for it.
Sometimes we simplify the trivially refined type
\tref{x}{\tclass}{\texttt{true}} to just \tclass. 

A \textit{program} is a sequence of method definitions and type annotations over methods. 
A method definition 
\pgdef
	{\methname{\tclass}{m}}
	{{\tref{x_1}{\tclass_1}{\expr_1}}, \dots, {\tref{x_n}{\tclass_n}{\expr_n}}}
	{\rtyp}
	{\tlabel}
	{\expr}
defines the method 
{\methname{\tclass}{m}}
with arguments $x_1,\dots,x_n$
and body $\expr$.
The type specification of the definition is a dependent function type: 
each argument binder $x_i$ can appear inside the 
arguments' refinements types $\expr_j$ for all $1\leq j \leq n$, 
and can also appear in the refinement of the result type $\rtyp$.
A method type annotation 
\typdef{\methname{\tclass}{m}}{\tfun{\overline{{\rtyp}}}{\rtyp}}{\tlabel} 
binds the method named $\methname{\tclass}{m}$
with the dependent function type
$\tfun{\overline{{\rtyp}}}{\rtyp}$. 
\corelan includes both method annotations
and  method definitions because annotations are used when a method's
code is not available, \eg in the cases of library methods, mixins, or metaprogramming. 

A \textit{label} \tlabel can appear in 
both method definitions and annotations 
to direct the method's translation into \rosette
as described in \S~\ref{subsec:method-calls}. 
The label \lexact 
states that a called method will be exactly translated
by using the translation of the body of the method. 
Since method type annotations do not have a body, 
they cannot be assigned the \lexact label. 
The \lpure label indicates that a method is pure 
and thus can be translated using an uninterpreted function.
Finally, the \lmodifies{\overline{\efieldread{\field}{x}}} label is
used when a method is impure, \ie it may modify its inputs. As we
saw earlier, the list $\overline{\efieldread{\field}{x}}$ captures all
the argument fields which the method may modify.
%%\NV{There is too much information about the translation here while we are only supposed to }
%%\NV{describe the language.}
%%\NV{These two should get decoupled, for clarity of presentation.}

\subsubsection{\coreinter.}
Figure~\ref{fig:inter-syntax} defines \coreinter, 
a core verification-oriented language that easily translates to \rosette.
% This is about implementation, so irrelevant here
% similar to the intermediate language we use for verification. 
% 
\corelan methods map to \coreinter functions, and 
\corelan objects map to a special \object struct type.
\coreinter provides primitives for creating, altering, and referencing instances of this type.
\textit{Values} in \coreinter consist of 
\textit{constants c} (defined identically to \corelan) and 
\vobject{i_1}{i_2}{\ibind{f_1}{\ival_1}\dots\ibind{f_n}{\ival_n}}, 
an instantiation of an \object type with 
class ID $i_1$, object ID $i_2$, and where each field $f_i$ 
of the \object is bound to value $\ival_i$. 
\textit{Expressions} include \letname bindings
(\elet{x}{\overline{\ibind{x_i}{\iexpr_i}}}{\iexpr}) where each $x_i$
may appear free in $\iexpr_j$ if $i<j$. They also include function calls, \assert,
\assume, and \return statements, as well as
\ehavoc{\efieldread{\field}{x}}, which mutates $x$'s field \field to a
fresh symbolic value. Finally, they include field assignment
\efieldass{x}{\field}{\iexpr} and field reads \efieldread{\field}{x}.

A \textit{program} is a series of definitions and verification queries. 
A \textit{definition} is a function definition
%an object definition \defineobj{x}{i_1}{i_2}{\ibind{\field_1}{u_1}\dots\ibind{\field_n}{u_n}}, 
%where \textit{x} gets bound to a new \object with class ID $i_1$, 
%object ID $i_2$, and field $\field_i$ bound to $u_i$, 
or a symbolic definition \definesymobj{x}{\ityp}, which binds
\textit{x} to either a fresh symbolic value if $\ityp$ is a solvable type (e.g.,
boolean, integer; see \S~\ref{subsec:overview:rosette}) or 
a new \object with
symbolic fields defined depending on the type of \ityp.
Finally, a verification query 
$\verify{\overline{\iexpr}}{\iexpr}$
checks the validity of \iexpr 
assuming $\overline{\iexpr}$.

\subsection{From \corelan to \coreinter}\label{subsec:tranlation}
\begin{figure}[p!]
\centering
\judgementHead{Expression Translation}{\translate{\statef}{\statem}{\expr}{\statef}{\statem}{\iexpr}}
$$
\inference{}{
\translate{\statef}{\statem}{c}{\statef}{\statem}{c}
}[\rtconst]
\quad
\inference{}{
\translate{\statef}{\statem}{x}{\statef}{\statem}{x}
}[\rtvar]
\quad
\inference{
	\translate{\statef}{\statem}{\expr_1}{\statef_1}{\statem_1}{\iexpr_1} &&
	\translate{\statef_1}{\statem_1}{\expr_2}{\statef_2}{\statem_2}{\iexpr_2}
}{
	\translate
	 {\statef}{\statem}{\eseq{\expr_1}{\expr_2}}
	 {\statef_2}{\statem_2}{\eseq{\iexpr_1}{\iexpr_2}}
}[\rtseq]
$$
\trspace
$$
\inference{
	\translate{\statef}{\statem}{\expr_1}{\statef_1}{\statem_1}{\iexpr_1}     &&
	\translate{\statef_1}{\statem_1}{\expr_2}{\statef_2}{\statem_2}{\iexpr_2} &&
	\translate{\statef_2}{\statem_2}{\expr_3}{\statef_3}{\statem_3}{\iexpr_3} 
}{
	\translate{\statef}{\statem}{\eif{\expr_1}{\expr_2}{\expr_3}}
	          {\statef_3}{\statem_3}{\eif{\iexpr_1}{\iexpr_2}{\iexpr_3}}
}[\rtif]
\ \ \
\inference{}{
	\translate
		{\statef}{\statem}{\eself}
		{\statef}{\statem}{\iself}
}[\rtself]
$$
\trspace
$$
\inference{
	\translate
		{\statef}{\statem}{\expr}
		{\statef'}{\statem'}{\iexpr}}{
\translate{\statef}{\statem}{\eass{x}{\expr}}{\statef}{\statem}{\eass{x}{u}}
}[\rtvass]
\qquad
\inference{
	\translate
		{\statef}{\statem}{\expr}
		{\statef'}{\statem'}{\iexpr}
}{
	\translate
		{\statef}{\statem}{\ereturn{\expr}}
		{\statef'}{\statem'}{\ereturn{\iexpr}}
}[\rtret]
$$
\trspace
$$
\inference{
	\extendfst{\field}{\statef}
}{
	\translate
	  {\statef}{\statem}{\einst{\field}}
	  {(\extendfst{\field}{\statef})}{\statem}{\efieldread{f}{\iself}}
}[\rtinst]
\quad
\inference{	
	\extendfst{\field}{\statef} && 
	\translate
		{\statef}{\statem}{\expr}
		{\statef'}{\statem'}{\iexpr}
}{
	\translate
		{\statef}{\statem}{\eass{\field}{\expr}}
		{(\extendfst{\field}{\statef})}{\statem}{\efieldass{\iself}{\field}{\iexpr}}
}[\rtiass]
$$
\trspace
$$
\inference{
	\classID{A} = i_c &&
	\freshObjectID{i_o} &&
    \extendfst{\field_i}{\statef} &&
}{
	\translate
		{\statef_0}
		{\statem_0}
		{\enew{\tclass}}
		{\statef_n}
		{\statem_n}
		{\vobject{i_c}{i_o}{\ibind{f_1}{\enil} \dots \ibind{f_{\sizef}}{\enil}}}
%		{\vobject{i_c}{i_o}{\overline{\ibind{f_i}{\enil}}}}
}[\rtnew]
$$
\trspace
$$
\inference{
	\rdlTypeOf{\expr_F} = \tclass &&
	\lexact =  \labelOf{\methname{A}{m}} \\
	\extendmst{\tclass\_m}{\statem} && 
	\translate{\statef}{\statem}{\expr_F}{\statef_1}{\statem_1}{\iexpr_F} &&
	\translate{\statef_i}{\statem_i}{\expr_i}{\statef_{i+1}}{\statem_{i+1}}{\iexpr_i}
        }{
	\translate
		{\statef}
		{\statem}
		{\ecall{\expr_F}{m}{\overline{\expr}}}
		{\statef_{n+1}}
		{(\extendmst{\tclass\_m}{\statem_{n+1}})}
		{\efunc{\tclass\_m}{\iexpr_F, \overline{\iexpr}}}
}[\rtnonmodcall]
$$
\trspace
$$
\inference{
	\rdlTypeOf{\expr_F} = \tclass &&
	\lpure =  \labelOf{\methname{A}{m}}\\
	\extendpst{\tclass\_m}{\statep} &&
    \freshVar{x, r}\\ 
	\typeOf{\methname{A}{m}} = 
	\tfun{{\tref{x}{\tclass_x}{\expr_x}}}{\tref{r}{\tclass_r}{\expr_r}} \\
	\translate{}{}{\expr_F}{}{}{\iexpr_F} &&
	\translate{}{}{\expr}{}{}{\iexpr} && 
	\translate{}{}{\expr_x}{}{}{\iexpr_x} &&
	\translate{}{}{\expr_r}{}{}{\iexpr_r}  
}{
	\translate{}{}
	 	{\ecall{\expr_F}{m}{\expr}}
	 	{}{}
    		{\begin{array}{l}
    		\elet{a}{\ibind{x}{u}
    		             \ibind{r}{\efunc{\tclass\_m}{\iexpr_F, a}}
    		            }{}
        \eseq{\eassert{\iexpr_x}}
    		{\eseq{\eassume{\iexpr_r}}
     	{r}}}
    		\end{array}
}[\rtmodpurecall]
$$
\trspace
$$
\inference{
	\rdlTypeOf{\expr_F} = \tclass &&
    \lmodifies{p} = \labelOf{\methname{A}{m}} \\
	\typeOf{\methname{A}{m}} = 
	\tfun{{\tref{x}{\tclass_x}{\expr_x}}}{\tref{r}{\tclass_r}{\expr_r}} \\ 
    h_x =  \{ u.f   \ |\ f\in \statef,\  x.f \in p \} && 
    h_F =  \{ u_F.f \ |\ f\in \statef,\  \eself.f \in p \}
    \\
    \freshVar{x, r} &&
	\translate{}{}{\expr_F}{}{}{\iexpr_F} &&
	\translate{}{}{\expr}{}{}{\iexpr} && 
	\translate{}{}{\expr_x}{}{}{\iexpr_x} &&
	\translate{}{}{\expr_r}{}{}{\iexpr_r}  
}{
	\translate
		{\statef}
		{\statem}
		{\ecall{\expr_F}{m}{\expr}}
		{\statef_4}
		{\statem_4}
    	  {\begin{array}{l}
    	  \elet{a_i}{\ibind{x}{u}}{\definesymobj{r}{\tclass_r};}\\
      \eseq{\eassert{\iexpr_x}}{
      \eseq{\ehavoc{h_F \cup h_x}}{\eseq{\eassume{\iexpr_r}}{r}}}
    	  \end{array}}
}[\rtmodimpurecall]
$$
\trspace
\trspace
\judgementHead{Program Translation}{\translate{\statef}{\statem}{\pg}{\statef}{\statem}{\ipg}}
$$
\inference{
}{
	\translate{}{}{\cdot}{}{}{\cdot}
}[\rtempty]
\quad
\inference{
	\translate{}{}{\pg}{}{}{\ipg}
}{
	\translate{}{}{\typdef{\methname{\tclass}{m}}{\tfun{{x_1}\text{:}{\rtyp_1}, \dots,{x_n}\text{:}{\rtyp_n}}{\rtyp}}{\tlabel}, \pg}
	          {}{}{\ipg}
}[\rtannot]
$$
\trspace
$$
\inference{
	\rtyp_i = \tref{x_i}{\tclass_{x_i}}{\expr_{x_i}} && 
	\rtyp = \tref{r}{\tclass_r}{\expr_r} \\
	\translate{\statef}{\statem}{\expr}{\statef_1}{\statem_1}{\iexpr} &&
	\translate{\statef_i}{\statem_i}{\expr_{x_i}}{\statef_{i+1}}{\statem_{i+1}}{\iexpr_{x_i}} &&
	\translate{\statef_{n+1}}{\statem_{n+1}}{\expr_r}
		      {\statef_{n+2}}{\statem_{n+2}}{\iexpr_r} &&
    \translate{}{}{\pg}{}{}{\ipg} && 1 \leq i\leq n
}{
	\translate
		{\statef}
		{\statem}
		{\pgdef{\methname{\tclass}{m}}
        		{{\rtyp_1}, \dots,{\rtyp_n}}
        		{\rtyp}
        {\tlabel}
        {\expr}, \pg}
		{\statef_2}
		{\statem_2}
		{\eseqfour{\define{A\_m}{\iself, x_1,\dots,x_n}{\iexpr};}
			{\definesymobj{self}{\tclass};}
		    {\definesymobj{x_i}{\tclass_{x_i}};}
		    {\eseq{\verify{\iexpr_{x_1},\dots,\iexpr_{x_n}}{\iexpr_r}}{\ipg}}
        }
}
[\rtdef]
$$
\caption{\textbf{Translation from \corelan to \coreinter}.
%We mix the notation $\overline{\expr_i}\equiv\overline{\expr}\equiv \expr_1, \dots, \expr_n$. 
For simplicity rules \rtmodpurecall and \rtmodimpurecall assume single argument methods.}
\label{fig:translation}
\end{figure}

Figure~\ref{fig:translation} defines the translation 
function 
$\translate{}{}{\expr}{}{}{\iexpr}$
that maps expressions (and programs) 
from \corelan to \coreinter.

\paragraph{Global States.}
The translation uses sets
\statem, \statep, and \statef,
to ensure all the 
methods, uninterpreted functions, and fields
are well-defined in the generated \coreinter term:
\begin{align*}
\statem &::= \methname{A_1}{m_1}, \dots, \methname{A_n}{m_n} &
\statep &::= \methname{A_1}{m_1}, \dots, \methname{A_n}{m_n} &
\statef &::= \field_1, \dots, \field_n
\end{align*}
In the translation rules, we use the standard set operations 
\extendmst{x}{\mathcal{X}} and $\mid \mathcal{X} \mid$ 
to check membership and size of the set $\mathcal{X}$.
Thus, the translation relation is defined over these sets: 
$\statem, \statep, \statef\vdash\translate{}{}{\expr}{}{}{\iexpr}$. 
Since the rules do not modify these environments, 
in Figure~\ref{fig:translation} we simplify the rules to 
$\translate{}{}{\expr}{}{}{\iexpr}$.
Note that even though the rules ``guess'' these environments by
making assumptions about which elements are members of the sets, 
in an algorithmic definition the rules can be used to construct the sets. 
% 

%%The inference rules for the function which translates \corelan to \coreinter 
%%are shown in Figure \ref{subsec:tranlation}. 
%%This function takes a triple $\langle\statef_1,\statem_1,s_1$, and returns a 
%%triple $\langle\statef_2,\statem_2,s_2$. $s_1$ is the actual structure from 
%%\corelan to be translated, and $s_2$ is the translated structure in \coreinter. 
%%%
%%$\statef_1$ is the set of all fields encountered prior to translating $s_1$, 
%%and $\statef_2$ is the set of all fields encountered after translating to $s_2$. 
%%%
%%The translation iterates until all the accumulated 
%%methods in the state \statem are translated.
%%Once finished, 
%%the final \statef is used to define the struct 
%%\object containing all fields in \statef
%%and the final \statep is used to defined all the required uninterpreted functions. 
%\statem is a set which accumulates all methods to be translated to \coreinter.

\paragraph{Expressions.}
The rules \rtconst and \rtvar are identity while 
the rules \rtif, \rtseq, \rtret, and \rtvass are trivially inductively defined. 
The rule \rtself translates \eself into the special \textit{variable} named \iself in \coreinter. 
The \iself variable is always in scope, since 
each \corelan method translates
to a \coreinter function with an explicit first argument named \iself.
The rules \rtinst and \rtiass translate a reference from and an assignment to the instance variable \einst{f}, 
to a read from and write to, respectively, the field \einst{f} of the variable \iself.
Moreover, both the rules assume the field \field to be in global field state \statef. 
The rule \rtnew translates from a constructor call \enew{\tclass} to an \object instance. The \classID{\tclass} function in the
premise of this rule returns the class ID of \tclass. 
The \freshObjectID{i_o} predicate ensures the new \object instance has a fresh object ID. 
Each field of the new \object, $f_1, \dots, f_{\sizef}$, is initially bound to \enil.

\paragraph{Method Calls.}
To translate the \corelan method call $\ecall{\expr_F}{m}{\overline{\expr}}$, 
we first use the function \rdlTypeOf{\expr_F} to 
type $\expr_F$ via \rdl type checking~\cite{Ren}. 
If $\expr_F$ is of type \tclass,
we split cases of the method call translation based on the 
value of \labelOf{\methname{A}{m}}, 
the label specified in the annotation of \methname{A}{m} 
(as informally described in~\S~\ref{subsec:method-calls}).

The rule \rtnonmodcall is used when the label is \lexact.
The receiver $\expr_F$ is translated to $\iexpr_F$ 
which becomes the first (\ie the \iself) argument of 
the function call to $A\_m$.
Moreover, 
\methname{A}{m} is assumed to be in the global method name set \statem
since it belongs to the transitive closure of the translation.

We note that for the sake of clarity, in the \rtmodpurecall
and \rtmodimpurecall rules, we assume that the method \methname{A}{m}
takes just one argument; the rules can be extended in a natural way
to account for more arguments.
The rule \rtmodpurecall is used when the label is \lpure.
In this case, the call is translated as an 
invocation to the uninterpreted function $A\_m$,
so \methname{A}{m} should be in the global set of uninterpreted functions \statep. 
The specification $\typeOf{\methname{A}{m}}$
of the method is also enforced.
Let $\tfun{{\tref{x}{\tclass_x}{\expr_x}}}{\tref{r}{\tclass_r}{\expr_r}}$
be the specification.
We assume that the binders in the specification are $\alpha$-renamed so that 
the binders $x$ and $r$ are fresh.
We use $x$ and $r$ to bind 
the argument and the result, respectively,
to ensure, via A-normal form conversion~\cite{Sabry}, 
that they will be evaluated exactly once, 
even though $x$ and $r$ may appear many times in the refinements.
To enforce the specification, we assert the method's 
precondition $e_x$ and assume the postcondition $e_r$.

If a method is labeled with \lmodifies{p} then the rule \rtmodimpurecall is applied. 
We locally define a new symbolic object as the return value, and we
\havoc the fields of all arguments (including \iself) specified in the \lmodifiesname label, 
thereby assigning these fields to new symbolic values.
Since we do not translate the called method at all,
no global state assumptions are made. 

\paragraph{Programs.}
Finally, we use the translation relation to translate programs from \corelan to \coreinter, \ie
\translate{}{}{\pg}{}{}{\ipg}.
The rule \rtannot discards type annotations.
The rule \rtdef translates a method definition for \methname{A}{m} to the function definition 
$A\_m$ that takes the additional first argument \iself. 
The rule also considers the declared type of \methname{A}{m} 
and instantiates a symbolic value for every input argument. 
Finally, all refinements from the inputs and output of the method type are translated 
and the derived verification query is made.

\subsection{From \coreinter to \rosette}\label{subsec:rosette}
We write $\torosette{\ipg}{R}$ to encode the translation 
of the \coreinter program $\ipg$ to the \rosette program $R$.
This translation is straightforward, since 
\coreinter consists of \rosette extended with some macros
to encode \ruby-verification specific operators, like 
\definesymname and \texttt{return}. 
In fact, in the implementation of the translation (\S~\ref{sec:evaluation}), we used  
Racket's macro expansion system to achieve this final transformation. 

\paragraph{Handling objects.} \coreinter contains multiple constructs for defining and altering objects, which are expanded
in Rosette to perform the associated operations over \object
structs. 
%\defineobj{x}{i_c}{i_o}{\ibind{f_1}{\iexpr_1}\dots\ibind{f_n}{\iexpr_n}} is a macro that binds
The expressions \vobject{i_c}{i_o}{\overline{\ibind{\field}{\ival}}} and \ehavoc{\efieldread{\field}{x}}, and the definition \definesymobj{x}{\ityp},
all described in \S~\ref{subsec:syntax}, are expanded to perform the corresponding
operations over values of the \object struct type.

\paragraph{Control Flow.}
Macro expansion is used to translate \texttt{return} and \texttt{assume} statements,
and exceptions into \rosette, since those forms are not built-in to the language.
To encode \return, we expand every function definition in \coreinter 
to keep track of a local variable \texttt{ret}, 
which is initialized to a special \texttt{undefined} value
and returned at the end of the function. 
We transform every statement \ereturn{e} to update the value of \texttt{ret} to \textit{e}. 
Then, we expand every expression \iexpr in a function to \unlessdone{\iexpr},
which checks the value of \texttt{ret}, proceeding with \iexpr if 
\texttt{ret} is \texttt{undefined} or skipping \iexpr if there is a
return value.

We used the encoding of \texttt{return} to encode more operators. 
For example, \texttt{assume} is encoded in \rosette as a macro 
that returns a special \texttt{fail} value when assumptions do not hold.
The verification query then needs to be updated with the condition that
\texttt{fail} is not returned.
A similar expansion is used
to encode and propagate exceptions.  

\subsection{Primitive Types}\label{subsec:primitives}

\corelan provides constructs for 
functions, assignments, control flow, etc, 
but does not provide the theories required to encode 
interesting verification 
properties that, for example, reason about 
booleans and numbers. 
On the other hand, \rosette is a verification oriented language
with special support for common theories over built-in
datatypes, including booleans, numeric types, and vectors. 
To bridge this gap, 
we encode certain \ruby expressions, 
such as constants $c$ in \corelan,
into \rosette's corresponding built-in datatypes. 

\paragraph{Equality and Booleans.}
To precisely reason about equality, 
we encode \ruby's \texttt{==} method over arbitrary objects 
using the object class' \textit{==} method if one is
defined. If the class inherits this method from Ruby's
top class, \textit{BasicObject}, then we encode \textit{==}
using \rosette's equality operator \texttt{equal?}
to check equality of object IDs.
We encode \ruby's booleans and operations over them
as \rosette's respective booleans and their operators.

\paragraph{Integers and Floats.}
By default, we encode \ruby's infinite-precision \rubinteger and \rubfloat objects
as \rosette's built-in infinite-precision \rosinteger and \rosreal
datatypes, respectively.
The infinite-precision encoding 
is efficient and precise, but it may result in undecidable queries
involving non-linear arithmetic or loops.
To perform (bounded) verification in such cases, we provide, via a configuration flag, the
option of
encoding \ruby's integers as \rosette's built-in finite sized
bitvectors.%, which come equipped with arithmetic operators. 

\paragraph{Arrays.}
Finally, we provide a special encoding for \ruby's arrays, which
are commonly used both for storing arbitrarily large random-access data
and to represent mixed-type tuples, stacks, queues, etc. 
We encode \ruby's arrays as a \rosette struct composed of a fixed-size
vector and an integer representing
the current size of the \ruby array.  Because we used fixed-size
vectors, we can only perform bounded verification over arrays.
% This is useful,
%however, since iterators over arrays represent one of the more common
%uses of loops in Ruby. \jeff{I don't understand the connection between
%  fixed-size and iterators.}
On the other hand, we avoid the need for loop invariants for iterators
and reasoning over array operations can be more efficient.
%%\begin{itemize}
%%\item Arrays need to resize dynamically when insertions occur past the end
%%\item Just as loop iterations need to be bounded for the solver to terminate, so does the size of Arrays need a bound. This makes vectors with a tweakable max size parameter a natural choice.
%%\item Rosette's vector operations index off of integers, making a standard integer an efficient choice for representing the size of a Ruby Array in the native translation code.
%%\end{itemize}

\subsection{Verification of \corelan}\label{subsec:soundness}

We define a verification algorithm \corerrdl that, 
given a \corelan program $\pg$,
checks if all the definitions satisfy their specifications. 
The pseudo-code for this algorithm is shown below:
\begin{rcode}
  def (*$\corerrdl(\pg)$*)
    ((*$\statef$*), (*$\statep$*), (*$\statem$*)) := guess((*$\pg$*))
    for ((*$f\in\statef$*)): add field (*$f$*) to object struct
    for ((*$u\in\statep$*)): define uninterpreted function (*$u$*)
    (*$\translate{}{}{\pg}{}{}{\ipg}  \torosette{}{R}$*)
    return if ((*\textit{valid}(R)*)) then SAFE else UNSAFE
  end
\end{rcode}
First, 
we ^guess^ the proper translation environments. 
In practice (as discussed in~\S~\ref{subsec:tranlation}), 
we use the translation of $\pg$ to generate the minimum 
environments for which translation of $\pg$ succeeds.
We define an \object struct in Rosette containing one
field for each member of \statef, and we define an uninterpreted
function for each method in \statep.
Next, we translate $\pg$ to a \coreinter program $\ipg$
via \translate{}{}{\pg}{}{}{\ipg} (\S~\ref{subsec:tranlation})
and $\ipg$ to a the \rosette program $R$,
via $\torosette{\ipg}{R}$ (\S~\ref{subsec:rosette}).
Finally, we run the \rosette program $R$.
The initial program $\pg$ is \textit{safe}, 
\ie no refinement type specifications are violated,
if and only if
the \rosette program $R$ is \textit{valid}, \ie all the 
\texttt{verify} queries are valid.

We conclude this section with a discussion of the \corerrdl verifier. 

\paragraph{\corerrdl is Partial.}
There exist expressions of \corelan that fail to translate into a \coreinter expression. 
The translation requires 
at each method call $\ecall{\expr_F}{m}{\overline{\expr}}$ that
the receiver has a class type \tclass.
There are two cases where this requirement fails:
(1) $\expr_F$ has a union type or
(2) type checking fails and so $\expr_F$ has no type. 
In our implementation (\S~\ref{sec:evaluation}), we extend the
translation to handle the first two cases. Handling for (1) is outlined in~\S~\ref{subsec:method-calls}.
Case (2) can be caused by either a type error in the program 
or a lack of typing information for the type checker. 
Translation cannot proceed in either case.

\paragraph{\corerrdl may Diverge.}
The translation to \rosette always terminates. 
All translation rules are inductively defined:
they only recurse on syntactically smaller expressions or programs.
Also, since the input program is finite, 
the minimum global environments required for translation are also finite. 
Finally, all the helper functions (including the type checking \rdlTypeOf{\cdot})
do terminate. 

Yet, verification may diverge, as the execution of the \rosette program may diverge. 
Specifications can encode arbitrary expressions, 
thus it is possible to encode undecidable verification queries. 
Consider, for instance, the following contrived \rosette program in which we 
attempt to verify an assertion over a recursive method:
\begin{rscode}
  (define (rec x) (rec x))
  (define-symbolic b boolean?)
  (verify (rec b))
\end{rscode}
\rosette attempts to symbolically evaluate this program, and thus diverges.

\paragraph{\corerrdl is Incomplete.}
Verification is incomplete and its precision relies 
on the precision of the specifications. 
For instance, 
if a pure method \methname{A}{m} is marked as impure, 
the verifier will not prove the congruence axiom.

%% Let's not talk about recursion since we do not explicitely encode it...
%%\rrdl is not complete, as there are valid specifications which 
%%we may not be able to verify. For instance, \rrdl may diverge when trying to
%%prove a postcondition on a recursive method.

%
\paragraph{\corerrdl is Sound.}
If the verifier decides that the input program is safe, 
then all definitions satisfy their specifications, 
assuming that 
(1) all the refinements are pure boolean expressions and
(2) all the labels are sound (i.e., methods match the specifications implied by the labels).
The assumption (1) is required since 
verification under diverging (let alone effectful)
specifications is difficult~\cite{halo}.
The assumption (2) is required since our translation encodes 
pure methods as uninterpreted functions,  
while for the impure methods it havocs only the unprotected 
arguments. 
%
%Checking these assumptions is left as future work. % (\S~\ref{sec:conclusion}).
%JF: We already said this was future work, no need to repeat.

\section{Evaluation}\label{sec:evaluation}
We implemented the 
\ruby refinement type checker \rrdl\footnote{Code available at: \url{https://github.com/mckaz/vmcai-rdl}} 
by extending \rdl~\cite{Ren} with refinement types. 
Table~\ref{eval_table} summarizes the evaluation of \rrdl. 

\paragraph{Benchmarks.} 
We evaluate RTR on six popular \ruby libraries:
\begin{itemize} % [leftmargin=*]
\item \money~\cite{Money} % \footnote{\code{https://github.com/RubyMoney/money}} 
performs currency conversions over monetary quantities and relies on mixin methods,
\item \businessTime~\cite{BusinessTime}  % \footnote{\url{https://github.com/bokmann/business_time}
performs time calculations in business hours and days,
\item \unitwise~\cite{Unitwise} % \footnote{\code{https://github.com/joshwlewis/unitwise}}
performs various unit conversions,
\item \geokit~\cite{Geokit} %\footnote{\code{https://github.com/geokit/geokit}}
performs calculations over locations on Earth,
\item \boxroom~\cite{Boxroom} %\footnote{\code{https://github.com/mischa78/boxroom}}
  is a \rails app for sharing files in a web browser and uses metaprogramming, and
\item \matrixLib~\cite{Matrix} %\footnote{\code{https://github.com/mischa78/boxroom}}
  is a \ruby standard library for matrix operations.
\end{itemize}
For verification, we forked the original \ruby libraries
and provided manually written method specifications in the form of refinement types. The forked repositories are publicly available \cite{Libraries}.
Experiments were conducted on a machine with a 3 GHz Intel Core i7
processor and 16 GB of memory.

We chose these libraries because they combine
\ruby-specific features challenging for verification,
like metaprogramming and mixins, 
with arithmetic-heavy operations. 
In all libraries we verify both
(1) \textit{functional correctness of arithmetic operations}  
(\eg no division-by-zero, the absolute value of a number should not be negative)
and (2) \textit{data-specific arithmetic invariants} 
(\eg integers representing months should always be in the range 
from \texttt{1} to \texttt{12} and
a \texttt{data} value added to an aggregate should always fall between maintained
\texttt{@min} and \texttt{@max} fields). 
In the \matrixLib library, we verify a matrix multiplication method,
checking that multiplying a matrix with $r$ rows by a matrix with $c$
columns yields a matrix of size $r\times c$. Note this method makes
extensive use of array operations, since matrices are implemented as
an array of arrays.

\paragraph{Quantitative Evaluation.}

\begin{table}[htbp]
%\vspace{-4em}
\caption{%Evaluation of \rrdl.
\textbf{Method} gives the class and name of the method verified.
%``\#'' separator indicates an instance method, while ``.'' indicates a class method.
\textbf{Ru-LoC} and \textbf{Ro-LoC} give number of LoC for
a Ruby method and the translated \rosette program.
\textbf{Spec} is the number of method and variable type annotations we had to write.
\textbf{Verification Time} is the median and semi-interquartile range of the time 
in seconds over 11 runs. 
\textbf{App Total} rows list the totals for an app, without double counting the same specs.
%Experiments were conducted on a machine with a 3 GHz Intel Core i7
%processor and 16 GB of memory.
\vspace{-1em}}
\label{eval_table}
% \vspace{-1em}
\begin{center}
\begin{tabular}{|r|r|r|r|r|r|}
\hline
\textbf{Method} & \textbf{Ru-LoC} & \textbf{Ro-LoC} & \textbf{Spec} & \multicolumn{2}{|c|}{\textbf{Verification Time}} \\
%\hline
\cline{5-6}
 & & & & \textbf{Time(s)}&\multicolumn{1}{c|}{\textbf{SIQR}} \\
\hline
\multicolumn{6}{|l|}{\textbf{\textit{\money}}} \\
\hline
 Money::Arithmetic\#-@ & 7 & 29 & 4 & 5.69& 0.14 \\ 
\hline
Money::Arithmetic\#eql? & 11 & 40  & 3 & 5.74& 0.03\\ 
\hline
Money::Arithmetic\#positive? & 5 & 24  & 3 & 5.40 & 0.01 \\ 
\hline
Money::Arithmetic\#negative? & 5 & 24  & 2 & 5.42& 0.01\\ 
\hline
Money::Arithmetic\#abs & 5 & 30  & 4 & 5.49& 0.01\\ 
\hline
Money::Arithmetic\#zero? & 5 & 26  & 2 & 5.38& 0.02\\ 
\hline
Money::Arithmetic\#nonzero? & 5 & 24  & 2 & 5.43& 0.03\\ 
\hline
 \textbf{App Total} & \textbf{43} & \textbf{197} & \textbf{10} & \textbf{38.56} & \textbf{0.25} \\
\hline
\multicolumn{6}{|l|}{\textbf{\textit{\businessTime}}} \\
\hline
ParsedTime\#- & 10 & 58  & 8 & 6.28& 0.02\\ 
\hline
BusinessHours\#initialize & 5 & 26 & 2 & 5.36& 0.04 \\ 
\hline
BusinessHours\#non\_negative\_hours? & 5 & 26 & 2 & 5.4& 0.01\\ 
\hline
Date\#week & 7 & 32 & 2 & 5.53& 0.01\\ 
\hline
Date\#quarter & 5 & 28 & 2 & 5.47& 0.00 \\ 
\hline
Date\#fiscal\_month\_offset & 5 & 25  & 2 & 5.41& 0.02\\ 
\hline
Date\#fiscal\_year\_week & 7 & 33  & 2 & 5.53& 0.03\\ 
\hline
Date\#fiscal\_year\_month & 12 & 35 & 3 & 5.65& 0.02 \\ 
\hline
Date\#fiscal\_year\_quarter & 9 & 42  & 2 & 5.72& 0.03\\ 
\hline
Date\#fiscal\_year & 11 & 32  & 4 & 5.81& 0.03\\ 
\hline
 \textbf{App Total} & \textbf{76} & \textbf{337} & \textbf{24} & \textbf{56.15} & \textbf{0.20} \\
\hline 
\multicolumn{6}{|l|}{\textbf{\textit{\unitwise}}} \\
\hline
Unitwise::Functional.to\_cel & 4 & 25  & 2 & 5.42& 0.03\\ 
\hline
Unitwise::Functional.from\_cel & 4 & 25  & 2 & 5.44& 0.03\\ 
\hline
Unitwise::Functional.to\_degf & 4 & 22 & 1 & 5.41& 0.01\\ 
\hline
Unitwise::Functional.from\_degf & 4 & 27  & 2 & 5.44& 0.02\\ 
\hline
Unitwise::Functional.to\_degre & 4 & 27 & 2 & 5.44& 0.01 \\ 
\hline
Unitwise::Functional.from\_degre & 4 & 27 & 2 & 5.42& 0.01 \\ 
\hline
 \textbf{App Total}  & \textbf{24} & \textbf{153} & \textbf{6} & \textbf{32.55} & \textbf{0.11}\\
\hline
\multicolumn{6}{|l|}{\textbf{\textit{\geokit}}} \\
\hline
Geokit::Bounds\#center & 7 & 31  & 4 & 5.4& 0.02\\ 
\hline
Geokit::Bounds\#crosses\_meridian? & 7 & 35 & 6 & 5.59& 0.12\\ 
\hline
Geokit::Bounds\#== & 9 & 60  & 5 & 5.97& 0.13\\ 
\hline
Geokit::GeoLoc\#province & 5 & 26 & 2 & 5.52& 0.11\\ 
\hline
Geokit::GeoLoc\#success? & 5 & 26 & 2 & 5.51& 0.05\\ 
\hline
Geokit::Polygon\#contains? & 26 & 68 & 10 & 10.8& 0.07\\ 
\hline
 \textbf{App Total} & \textbf{59} & \textbf{246}  & \textbf{21} & \textbf{38.80} & \textbf{0.50}\\
\hline
\multicolumn{6}{|l|}{\textbf{\textit{\boxroom}}} \\
\hline
 UserFile\#move & 12 & 34 & 3 & 5.57& 0.05 \\ 
\hline
\multicolumn{6}{|l|}{\textbf{\textit{\matrixLib}}} \\
\hline
 Matrix.* & 57 & 94 & 9 & 334.35 & 3.99  \\
\specialrule{1pt}{1pt}{1pt}
\textbf{Total}  & \textbf{271} & \textbf{1061} & \textbf{73} & \textbf{505.98} & \textbf{5.10}  \\
\hline
\end{tabular}
\vspace{-3em}
\end{center}
\end{table}

Table~\ref{eval_table}
summarizes our evaluation quantitatively. 
For each application, we list every verified \textbf{Method}.
%
%\rrdl acts at a method-level granularity, 
%that is, 
%the programmer can specify exactly which methods should be verified,
%and when they should be verified: immediately when it is defined,
%prior to executing the method when it is called, or at a customized
%time.
%
%Yet, verified methods co-exist with (and can even call)
%the non-verified, modularly used methods. 
%
In our experiment, we focused on methods with interesting arithmetic properties.

The \textbf{\ruby LoC} column gives the size of the verified Ruby program. This metric
includes the lines of 
all methods and annotations that were used to verify the method in question.
For each verified method, \rrdl generates
a separate Rosette program. 
We give the sizes of these resulting programs in the \textbf{\rosette LoC} column.
Unsurprisingly, the LoC of the \rosette program
increases with the size of the source \ruby program. 

We present the median (\textbf{Time(s)}) and semi-interquartile range (\textbf{SIQR}) of the \textbf{Verification Time} 
required to verify all methods for an application over 11 runs. 
For each verified method, the \textbf{SIQR} was at most 2\% of the verification time, 
indicating relatively little variance in the verification time.
Overall, verification was fast, as might be expected for relatively small
methods. The one exception was matrix multiplication. In this case,
the slowdown was due to the extensive use of array operations
mentioned above. We bounded array size (see \S~\ref{subsec:primitives}) at 10
for the evaluations. For symbolic arrays, this means \rosette must reason about
every possible size of an array up to 10. This burden is exacerbated by 
matrix multiplication's use of two symbolic two-dimensional arrays. 

% We do note that the time taken to verify the \matrixLib multiplication
% method was significantly higher than methods verified in other
% libraries. In general, \rrdl takes longer to verify assertions
% involving heavy array operations. This is magnified by the
% \matrixLib library, which uses two-dimensional arrays.

Finally, Table~\ref{eval_table} lists the number of type 
\textbf{specifications} required to verify each method. 
These are comprised of method type annotations, including the 
refinement type annotations for the verified methods 
themselves, and variable type annotations for instance variables. 
Note that we do not quantify the number of type annotations used
for \ruby's core and standard libraries, since these are included in
\rdl. 
%%\jeff{Did those include any refinement types? If so we should mention that.}
%%They do not include any refinement types.

We observe that there is significant variation in the number of
annotations required for each application.  For example, \unitwise
required 6 annotations to verify 6 methods, while \geokit required 21
annotations for 6 methods. The differences are due to code
variations: To verify a method, the programmer needs to give a
refinement type for the method plus a type for each instance variable
used by the method and for each (non-standard/core library) method
called by the method.

\paragraph*{Case Study.}

Next we illustrate the \rrdl verification process by presenting the exact steps 
required to specify and check the properties of a method from 
an existing \ruby library.
For this example, we chose to verify the ^<<^ method of the 
\texttt{Aggregate} library~\cite{Aggregate}, 
a \ruby library for aggregating and performing statistical computations over some numeric data. 
The method ^<<^
takes one input, ^data^, and adds it to the aggregate by updating
(1) the minimum ^@min^ and maximum ^@max^ of the aggregate, 
(2) the count ^@count^, sum ^@sum^, and sum of squares ^@sum2^ of the aggregate, and finally 
(3) the correct bucket in ^@buckets^.

\begin{rcode}
  def <<(data)
    if 0 == @count
      @min = data ; @max = data
    else
      @max = data if data > @max ; @min = data if data < @min
    end
    @count += 1 ; @sum   += data ; @sum2  += (data * data)
    @buckets[to_index(data)] += 1 unless outlier?(data)
  end
\end{rcode}

We specify functional correctness of the method ^<<^
by providing a refinement type specification that declares that 
after the method is executed, 
the input ^data^ will fall between ^@min^ and ^@max^. 
\begin{rcode}
  type :<<,  `(Integer data) -> Integer { @min<=data<=@max }', verify: :bind
\end{rcode}

Here, the symbol ^:bind^ is an arbitrary label. To verify the
specification, we load the library and call the verifier with this label:
\begin{rcode}
  rdl_do_verify :bind 
\end{rcode}
RTR proceeds with verification in three steps:
\begin{itemize}
\item first use \rdl to type check the basic types of the method, 
\item then translate the method to \rosette (using the translation of \S~\ref{sec:formalism}), and
\item finally run the \rosette program to check the validity of the specification.
\end{itemize}

Initially, verification fails in the first step with the error
\begin{rscode}
  error: no type for instance variable `@count'
\end{rscode}
To fix this error, the user needs to provide the correct types for the instance variables
using the following type annotations.
\begin{rcode} % [style=ruby,firstnumber=18]
  var_type :@count, `Integer'
  var_type :@min, :@max, :@sum, :@sum2, `Float'
  var_type :@buckets, `Array<Integer>'
\end{rcode}
The ^<<^ method also calls two methods that are 
not from Ruby's standard and core libraries: 
^to_index^, which takes a numeric input 
and determines the index of the bucket the input falls in, 
and ^outlier?^, which determines if the given 
data is an outlier based on provided specifications from the programmer. 
These methods are challenging to verify.
For example, 
the ^to_index^ method makes use of non-linear arithmetic in the form of logarithms, 
and it includes a loop.
Yet, neither of the calls ^to_index^ or ^outlier?^ should affect verification of the specification of ^<<^.
So, it suffices to provide type annotations with a \lpure label, indicating we want to use uninterpreted functions to represent them:
\begin{rcode} % [style=ruby,firstnumber=24]
  type :outlier?, '(Float i) -> Bool b', :pure
  type :to_index, '(Float i) -> Integer out', :pure
\end{rcode}
Given these annotations, the verifier has enough information to prove the postcondition on 
^<<^, and it will return the following message to the user: 
\begin{rcode} % [style=ruby,firstnumber=24]
  Aggregate instance method << is safe.
\end{rcode}

When verification fails, an unsafe message is provided, 
combined with a counterexample consisting of bindings to symbolic values that causes the postcondition to fail. 
For instance, if the programmer \textit{incorrectly} specified 
that  ^data^ is less than the ^@min^, \ie
\begin{rcode}
  type :<<, `(Integer data) -> Integer { data (*$<$*) @min }'
\end{rcode}
Then \rrdl would return the following message:
\begin{rcode}
  Aggregate instance method << is unsafe. 
  Counterexample: (model [real_data 0][real_@min 0] (*\dots*))
\end{rcode}
This gives a binding to symbolic values in the translated \rosette program 
which would cause the specification to fail. 
We only show the bindings relevant to the specification here: 
when ^real_data^ and ^real_@min^, 
the symbolic values corresponding to ^data^ and ^@min^ respectively, 
are both 0, the specification fails.

\section{Related Work}\label{sec:related}

\paragraph{Verification for Ruby on Rails.}
Several prior systems can verify properties of Rails apps.
\textit{Space}~\cite{Near:2016:FSB:2884781.2884836} detects
security bugs in Rails apps by using symbolic execution to generate a
model of data exposures in the app and reporting a bug if the model
does not match common access control patterns.
Boci\'{c} and Bultan proposes \textit{symbolic model
  extraction}~\cite{Bocic:2017:SME:3097368.3097455}, which extracts
models from Rails apps at runtime, to handle metaprogramming. The
generated models are then used to verify data integrity and access
control properties.
\textit{Rubicon}~\cite{Near:2012:RBV:2393596.2393667} allows
programmers to write specifications using a domain-specific language
that looks similar to Rails tests, but with the ability to quantify
over objects, and then checks such specifications with bounded
verification. \textit{Rubyx}~\cite{Chaudhuri:2010:SSA:1866307.1866373}
likewise allows programmers to write their own specifications over
Rails apps and uses symbolic execution to verify these
specifications.

In contrast to RTR, all of these tools are specific to Rails and do
not apply to general Ruby programs, and the first two systems do not
allow programmers to specify their own properties to be verified.

\paragraph{Rosette.}

Rosette has been used to help establish the security and reliability
of several real-world software systems. \citet{Pernsteiner2016} use
Rosette to build a verifier to study the safety of the software on a
radiotherapy machine. \textit{Bagpipe}
\cite{Weitz:2016:SVB:2983990.2984012} builds a verifier using
Rosette to analyze the routing protocols used by Internet Service
Providers (ISPs). These results show that Rosette can be applied in a
variety of domains.

\paragraph{Types For Dynamic Languages.}

There have been a number of efforts to bring type systems to dynamic
languages including Python
\cite{Ancona:2007:RST:1297081.1297091,Aycock2000}, Racket
\cite{Tobin-Hochstadt2006,Tobin-Hochstadt2008}, and JavaScript
\cite{Anderson2005,Lerner2013,Thiemann2005}, among others. However,
these systems do not support refinement types.

Some systems have been developed to introduce refinement types to
scripting and dynamic languages. \textit{Refined TypeScript}
(RSC)~\cite{Vekris16} introduces refinement types to
TypeScript~\cite{Bierman2014,Rastogi2015}, a superset of JavaScript
that includes optional static typing. RSC uses the framework of Liquid Types
\cite{Rondon08} to achieve refinement inference. Refinement types have
been introduced \cite{Kent16} to Typed Racket as well. As far as we
are aware, these systems do not support mixins or metaprogramming.

% While both of
% these efforts introduce refinement types to dynamic languages, both
% attempt to check these types \textit{at compile time}. In contrast, by
% using the just-in-time approach, \rrdl is able to generate and verify
% expressive specifications in the face of highly dynamic features such
% as metaprogramming. Furthermore, by including method annotation labels
% describing the purity or side effects of a method, \rrdl allows for
% more expressive annotations to be used in modular verification.
% \MK{Mention mixins here as well? Unclear if other systems handle... no
%   mention of it.}
%

% \textit{Hybrid type checking} \cite{Flanagan2006} allows for arbitrary
% predicates in refinement types, and proposes statically verifying
% refinement types whenever possible, but inserting dynamic checks when
% static verification is not possible. This is in some ways similar to
% \rrdl, since we allow optionally treating refinement types as
% dynamically checked contracts. \MK{Do we mention this anywhere?}
% However, our approach of performing not just dynamic checks but also
% static verification \textit{at runtime} allows us to verify refinement
% types in the face of metaprogramming, and to achieve assume-guarantee
% reasoning in the partial environments of mixins.

\paragraph{General Purpose Verification}

Dafny~\cite{Leino10} is an object-oriented language with built-in
constructs for high-level specification and verification. While it
does not explicitly include refinement types, the ability to specify a
method's type and pre- and postconditions effectively achieves the
same level of expressiveness. Dafny also performs modular verification
by using a method's pre- and postconditions and labels 
indicating its
purity or arguments mutated, an approach RTR largely emulates.
However, unlike Dafny, \rrdl leaves this modular treatment of methods
as an option for the programmer. Furthermore, unlike RTR, Dafny does not include
features such as mixins and metaprogramming.

\section{Conclusion and Future Work}\label{sec:conclusion}

We formalized and implemented \rrdl,  
a refinement type checker for \ruby programs
using assume-guarantee reasoning and the just-in-time checking technique of \rdl. 
Verification at runtime
naturally adjusts standard refinement types to handle 
\ruby's dynamic features, such as metaprogramming and mixins. 
To evaluate our technique, we used \rrdl 
to verify numeric properties on six commonly used \ruby and Ruby on Rails
applications, by adding refinement type specifications
to the existing method definitions. We found that verifying these methods
took a reasonable runtime and annotation burden, and thus we believe
\rrdl is a promising first step towards bringing verification to \ruby.

Our work opens new directions for further \ruby verification. 
We plan to explore verification of purity and immutability labels,
which are currently trusted by RTR. We also plan to develop refinement
type inference by adapting Hindley-Milner and liquid typing~\cite{Rondon08}
to the RDL setting, and by exploring whether \rosette's
synthesis constructs could be used for refinement inference.
We will also extend the expressiveness of RTR 
by adding support for loop invariants and dynamically defined instance variables,
among other \ruby constructs. Finally, as \ruby
is commonly used in the Ruby on Rails framework, we will extend
\rrdl with modeling for web-specific constructs such as access control protocols
and database operations to further support verification in the domain of web applications.

%%
%%We plan to continue work on \rrdl so that it can be applied to a larger class of Ruby programs 
%%and verify more expressive properties. 
%%Ruby is commonly used for web application development in frameworks like Ruby on Rails, 
%%so we plan to model more web-specific constructs such as access control protocols and database operations in order to verify properties over this domain. We also have more work to do to increase the expressiveness of \rrdl and handle more Ruby properties, such as dynamically 
%%
%% reflection on new objects
%% Check check labels
%% inference
%% loop invariants

\section*{Acknowledgements}

We thank Thomas Gilray and the anonymous reviewers for their feedback
on earlier versions of this paper.
This work is supported in part by NSF CCF-1319666, CCF-1518844, CCF-1618756, CCF-1651225, CNS-1518765, and DGE-1322106.

\bibliographystyle{splncsnat}
\bibliography{refs}

\end{document}